\documentstyle[11pt]{article}
\begin{document}

\title{Applied Symbolic Dynamics\protect\footnote{Based on an invited talk
at the 1998 Annual Meeting of the Physical Society held at NCU, Chung-li.}}
\author{Bai-lin HAO\thanks{On leave from the Institute of Theoretical Physics,
P. O. Box 2735, Beijing 100080.}
\\Center for Complex Systems\\National Central University, Chung-li,
Taiwan 320\\and\\
National Center for Theoretical Sciences\\P. O. Box 2-131, Hsinchu, Taiwan 300}
\date{NSC-CTS-980502 (10 May 1998)}
\maketitle

\begin{abstract} 
Symbolic dynamics is a coarse-grained description of dynamics. By taking
into account the ``geometry'' of the dynamics, it can be cast into a powerful
tool for practitioners in nonlinear science. Detailed symbolic dynamics can be
developed not only for one-dimensional mappings, unimodal as well as those
with multiple critical points and discontinuities, but also for some
two-dimensional mappings. The latter paves the way for symbolic dynamics study
of ordinary differential equations via the Poincar\'e maps. This paper provides
an overview of the recent development of the applied aspects of symbolic
dynamics.
\end{abstract}

\bigskip


\bigskip

\section*{I. Introduction}

Symbolic dynamics is a {\it rigorous} way to study complex dynamics with
{\it finite} precision. As an abstract chapter in the mathematical theory
of dynamical systems${}^{\cite{mh,ay}}$, it originated from the work of
Hadamard${}^{\cite{ad}}$ and Morse${}^{\cite{morse}}$. The basic idea is very
simple: divide the phase space into a finite number of regions and label each
region by a letter from a certain alphabet; instead of following a trajectory
point by point one only keeps recording
the alternation of letters. One loses a great amount of detailed information
on the dynamics, but some essential, robust, features of the motion may be
kept, e.g., periodicity or chaoticity of an orbit. This is nothing but what
physicists call a coarse-grained description.

The idea of symbolic dynamics applies to dynamics in any finite-dimensional
phase space. In many cases, say, for theorem-proving, an arbitrary partition of
the phase space would do the job. However, only for one-dimensional mappings
symbolic dynamics has been developed more or less completely. This is due to
the nice ordering property of real numbers on an interval and due to the
possibility of partitioning the ``phase space'', i.e., the interval, in
accordance with the ``geometry'' of the dynamics. In fact, many useful rules
and beautiful results have been derived. Recently, significant progress has
been made in symbolic dynamics of two-dimensional maps, but the achievement is
still rather limited compared to what has been known in one dimension.
Nevertheless, the knowledge of symbolic dynamics in one and two dimensions
proves to be quite instructive in understanding the systematics of periodic
orbits and chaotic behavior in some higher-dimensional dissipative systems,
e.g., the Lorenz model and some periodically forced nonlinear oscillators.
The presence of dissipation is essential, since it causes the shrinking of
phase space volume, which makes a ``strange attractor'' closer to a
low-dimensional objects at least in certain sections of the attractor.

Chaotic dynamics of dissipative systems provides a rare and lucky case in
physics, when low-dimensional systems are not merely toy models, but lead to
essential ``universal'' results which are quite useful in understanding
higher dimensional systems. In a sense, everyone who enters the field of chaos
should start with the study of symbolic dynamics. We have called this
approach {\it applied symbolic dynamics}${}^{\cite{h89,zh,zh94,hz98}}$.

Applied symbolic dynamics commenced from a seminal paper by Metropolis, Stein
and Stein${}^{\cite{mss}}$. The kneading theory of Milnor and
Thurston${}^{\cite{mt}}$, the lecture of Guckenheimer${}^{\cite{guck}}$, and a
paper by Derrida, Gervois, and Pomeau${}^{\cite{dgp}}$, among others, further
developed the theory. What had been known by the end of 1970s was summarized in
the book by Collet and Eckmann${}{\cite{ce80}}$. There has been significant
generalization and simplification of the theory both for one-dimensional and
two-dimensional mappings as well as their application to ordinary
differential equations since the mid 1980s, for details see, e.g., 
\cite{zh94,hz98}.

\section*{II. One-Dimensional Mappings}

We consider one-dimensional maps of the general form
$$ x_{n+1}=f(\mu, x_n), $$
where $f(\mu, x)$ is a nonlinear ``mapping function'' of the variable $x$ and
$\mu$ is a set of parameters. The function $f(x)$ maps an interval $I$ into
itself; it may have several monotone pieces between ``turning'' points and 
discontinuities. Symbolic dynamics of such maps has been understood more or
less completely. We summarize some main points:

\indent 1. The phase space, i.e., the interval, is partitioned according to
 the monotone branches of the mapping function. Any numerical orbit corresponds
 to a semi-infinite symbolic sequence and a functional composition
 represented by the same set of symbols, understood as inverse functions of
 the monotone branches.\\
\indent 2. All symbolic sequences for a given type of maps may be ordered.
 Admissibility conditions based on ordering rules may be formulated to test
 whether a given symbolic sequence is reproducible in the dynamics or not.\\
\indent 3. There is a Periodic Window Theorem: any admissible superstable
 periodic sequence may be extended to a ``window'' with its upper and lower
 sequences. It leads to a method of generating the shortest admissible
 superstable periodic sequence in between any two given admissible
 periodic sequences.\\
\indent 4. There is a word-lifting technique${}^{\cite{zhwc84,hz89}}$ which
 allows one to determine the parameter of any given type of superstable
 periodic and eventually periodic orbit.\\
\indent 5. There are composition rules which generate more admissible sequences
 from known ones. The simplest rule is called
 the $\ast$-composition${}^{\cite{dgp}}$ and it has a close relation with
 possible fine structure in the power spectra of observed periodic orbits.\\
\indent 6. The counting problem on how many periodic orbits exist for
 a given map has been solved completely for continuous maps${}^{\cite{xh94}}$
 and partly for maps with discontinuity${}^{\cite{xh95}}$.
\indent 7. Topological entropy of superstable periodic and eventually periodic
 sequences may be calculated from transfer matrices which may be written
 down directly from the symbolic sequences without knowing the precise form
 of the mapping function.\\
\indent 8. Maps with multiple critical points and discontinuities are best
 parameterized by their kneading sequences. The parameter space, called also
 the kneading space, may be constructed by using the admissibility
 conditions.\\
\indent 9. Circle maps, i.e., maps from a circle to itself, though may be
 studied as that with multiple critical points and discontinuities, do
 possess some specific features dictated by the topology of the phase space.
 Their study is facilitated by the Farey representation of rational
 numbers and the associated machinery, see, e.g., Chapter~4 of \cite{hz98}.\\
\indent 10. For maps with a discrete symmetry there are the phenomena of
 symmetry breaking and restoration which may be analyzed by using symbolic
 dynamics${}^{\cite{zh89}}$.\\
\indent 11. Symbolic sequences in the unimodal maps are naturally related to
 formal language and the theory of grammatical complexity. Periodic and
 eventually periodic sequences are the only types of regular language. The
 transfer matrix provides a way to go beyond regular
 languages${}^{\cite{h91}}$. There
 are examples of context-dependent languages of different complexity but
 no known example of context-free language yet. Hence a conjecture: no
 context-free language exists in the languages associated with unimodal maps.
 A good reference to this set of problems is \cite{xie96}.\\
\indent 12. Periodic orbits in unimodal maps are related to knots in 3-space.
 There are some observations but not much rigorous results, see, e.g.,
 Chapter~9 in \cite{hz98}.

\section*{III. Two-Dimensional Mappings}

In two- and higher-dimensional systems the nice ordering property of real
numbers and the simple partition of an interval, which have played crucial role
in symbolic dynamics of one-dimensional maps, no longer exist. In addition,
2D maps usually lead to bi-infinite symbolic sequences. The partition
for the H\'non map${}^{\cite{henon}}$, using tangencies of the invariant
manifolds of the fixed points, was first discussed by Grassberger and
Kantz${}^{\cite{gk85}}$ in order to calculate its topological entropy. Then
Cvitanovi\'c, Gunaratne, and Procaccia${}^{\cite{cgp}}$ used the partition
to develop symbolic dynamics. Later on the role of forward and backward
foliations of the map in determining the partition lines has been recognized
by Zheng and collaborators. In fact, the generalization from tangent points
between the stable and unstable manifolds to that between the two dynamical
foliations are essential and necessary, as it was shown analytically on
the example of two piecewise linear maps.
The simplest case turns out to be the two-dimensional version of the sawtooth
map, introduced by T\'el${}^{\cite{tel}}$. Its symbolic dynamics was
constructed in \cite{z92}. The piecewise linear counterpart of the H\'enon map,
so-called Lozi map${}^{\cite{lozi}}$, may be treated in a similar
manner${}^{\cite{z9192,zl94}}$. The two piecewise linear maps helped to reach a
deeper understanding of the symbolic dynamics of the H\'enon
map${}^{\cite{zzg92,zz93,f94,z96}}$. However, the symbolic dynamics of
H\'enon map has not been understood thoroughly on the whole parameter plane.

We mention in passing that dynamical foliations may be constructed for
Hamiltonian systems as well especially when the Poincar\'e sections may be
reduced to two-dimensional, hence leading to a better symbolic dynamics,
see \cite{z97a,z97b}.

\section*{VI. Ordinary Differential Equations}

In order to cut a long story short we will only mention a few results
of applying symbolic dynamics to ordinary differential equations.

Some years ago we have applied symbolic dynamics of one-dimensional maps to
the systematics of periodic orbits in differential equations. In particular,
the ordering of periodic orbits of the periodically forced
Brusselator${}^{\cite{hz82}}$ has been compared to that of the quadratic map,
using symbolic dynamics of two letters${}^{\cite{hwz84}}$. The systematics of
periodic orbits${}^{\cite{dhh85}}$ in the autonomous Lorenz model has been
juxtaposed with the ordering of kneading sequences in the antisymmetric cubic
map with and without a discontinuity${}^{\cite{dh88}}$. These essentially
one-dimensional studies were summarized in \cite{h86}.

Our main argument for using 1D maps lies in the shrinking of phase space volume
due to dissipation. However, the Poincar\'e maps of ordinary differential
equations are necessarily two-dimensional and there is no {\it a priori}
reason that the two-dimensional nature will not show off. Having reached a
better understanding of symbolic dynamics of two-dimensional maps, we have
undertaken the job of justifying the previous one-dimensional approach and
revealing the cases where a two-dimensional study leads to essentially new
insight. We list some recent references: the periodically forced
Brusselator${}^{\cite{lz95,lzh96}}$, the forced two-well Duffing
equation${}^{\cite{xzh95}}$, the NMR-laser model${}^{\cite{zl95,lwz96}}$,
and the Lorenz model${}^{\cite{f95,fh96,zl97,hlz98}}$. In the Lorenz
equations numerical work under the guidance of topology, i.e., symbolic
dynamics, has yield all stable and unstable periodic orbits up to period~6
in a wide parameter range${}^{\cite{hlz98}}$. 

\section*{Acknowledgments}

The author acknowledges the hospitality and support of the National Central
University, Chung-li, and the National Center for Theoretical Sciences,
Hsinchu.

\begin{small}

\end{small}
\end{document}